\documentclass[12pt]{article}

\usepackage{axodraw}
\usepackage{epsf}
\usepackage{rotating}

\catcode`@=11
\def\citer{\@ifnextchar
[{\@tempswatrue\@citexr}{\@tempswafalse\@citexr[]}}

\def\@citexr[#1]#2{\if@filesw\immediate\write\@auxout{\string\citation{#2}}\fi
  \def\@citea{}\@cite{\@for\@citeb:=#2\do
    {\@citea\def\@citea{--\penalty\@m}\@ifundefined
       {b@\@citeb}{{\bf ?}\@warning
       {Citation `\@citeb' on page \thepage \space undefined}}%
\hbox{\csname b@\@citeb\endcsname}}}{#1}}
\catcode`@=12

\newcommand{\beq}{\begin{eqnarray}}
\newcommand{\eeq}{\end{eqnarray}}
\newcommand{\nn}{\noindent}

\newcommand{\gsim}{\raisebox{-0.13cm}{~\shortstack{$>$ \\[-0.07cm] $\sim$}}~}

\begin{document}

\thispagestyle{empty}

\def\thefootnote{\fnsymbol{footnote}}

\begin{flushright}
PSI--PR--03--07 \\
hep-ph/0305288
\end{flushright}

\vspace{1cm}

\begin{center}

{\large\sc A Note on Doubly-Charged Higgs Pair Production} \\[0.5cm]

{\large\sc at Hadron Colliders}%
\footnote{This work has been supported in part by the Swiss Bundesamt
f\"ur Bildung und Wissenschaft and by the European Union under contract
HPRN-CT-2000-00149.}

\vspace{1cm}

{\sc Margarete M\"uhlleitner and Michael Spira}

\vspace{0.5cm}

{\it Paul Scherrer Institut, CH-5232 Villigen PSI, Switzerland}

\end{center}

\vspace{2cm}

\begin{abstract}
\nn
We analyze the next-to-leading order QCD corrections to the production
of doubly-charged Higgs particles at hadron colliders in extensions of
the SM with Higgs isospin triplets. At both the Tevatron and the LHC,
these corrections are found to be moderate in size increasing the cross
sections by about 20--30\%. The residual theoretical uncertainties are
of the order of 10--15\% which is sufficient for experimental searches
for these particles at the Tevatron and LHC.
\end{abstract}

\newpage

\def\thefootnote{\arabic{footnote}}
\setcounter{footnote}{0}

\section{Introduction}
%        ============
Exotic extensions of the Higgs sector involving higher isospin
multiplets naturally predict the existence of doubly-charged Higgs
bosons $\Delta^{\pm\pm}$. Particular examples are left-right symmetric
models \cite{lrsym}. However, higher Higgs multiplets are generally
severely constrained by the $\rho$ parameter which is unity at
tree-level. In order to fulfill these constraints, very particular Higgs
representations have to be chosen or fine-tuning is required between
different Higgs multiplets. The simplest options allowed by the $\rho$
parameter are Higgs multiplets without neutral states or
representations containing neutral states with a very small vacuum
expectation value. Left-right symmetric models predict the appearance of
a left- and a right-handed Higgs triplet, both with hypercharge $|Y|=2$
\cite{lrsym}. If the vacuum expectation values of the neutral members
vanish, the doubly-charged components $\Delta^{\pm\pm}$ do not couple to
$W^\pm W^\pm$ pairs. In this case the dominant doubly-charged Higgs
production process at hadron colliders is pair production via $q\bar
q\to \gamma^*,Z^*\to \Delta^{++} \Delta^{--}$ \cite{gunion}. The cross
section of this production mode only depends on the electroweak quantum
numbers and the mass of the doubly-charged Higgs states and not on
further details of the model.  Doubly-charged Higgs bosons have been
searched for at the LEP collider via the related process $e^+e^-\to
\gamma^*,Z^*\to \Delta^{++} \Delta^{--}$ resulting in a lower mass
bound $M_\Delta\gsim 98.5$ GeV within supersymmetric left-right
symmetric models \cite{opal}. Present searches at the Tevatron cannot
impose any mass limits yet, but this will improve with increasing
statistics \cite{tevatron}.

\section{QCD corrections to the production processes}
%        ===========================================
At hadron colliders, the lowest order partonic cross section for
doubly-charged Higgs boson pair production is given by
\beq
\hat{\sigma}_{\rm LO}(q\bar{q} \to \Delta^{++} \Delta^{--}) =
\frac{\pi\alpha^2}{9Q^2} \beta^3 \left[ e_q^2 e_\Delta^2 + \frac{e_q
e_\Delta v_q v_\Delta(1- M_Z^2/Q^2) + (v_q^2+a_q^2)v_\Delta^2} {(1-
M_Z^2/Q^2)^2 + M_Z^2\Gamma_Z^2/Q^4} \right]
\eeq
with $v_q=(2I_{3q}-4e_q s_W^2)/(2s_Wc_W), a_q=2I_{3q}/(2s_Wc_W)$ and
$v_\Delta =(2I_{3\Delta}-2e_\Delta s_W^2)/(2s_Wc_W)$, where
$I_{3q}~(I_{3\Delta})$ denotes the third isospin component and
$e_q~(e_\Delta)$ the electric charge of the quark $q$ (doubly charged
Higgs boson $\Delta^{--}$) and $s_W=\sin\theta_W, c_W=\cos\theta_W$.
$Q^2$ is the squared partonic c.m.\,energy, $\alpha$ the QED coupling
evaluated at the scale $Q$, $M_Z$ the $Z$ boson mass and $\Gamma_Z$ the
$Z$ boson width. The Higgs velocity is defined as
$\beta=\sqrt{1-4M_\Delta^2/Q^2}$.

The hadronic cross sections can be obtained from convoluting the
partonic cross section with the corresponding (anti)quark densities of
the (anti)protons 
\begin{eqnarray}
\sigma_{LO} (p\raisebox{-0.00cm}{\shortstack{{\tiny (---)} \\[-0.15cm]
$p$}} \to \Delta^{++} \Delta^{--}) = \int_{\tau_0}^1 d\tau 
\sum_q \frac{d{\cal L}^{q\bar q}}{d\tau} \hat\sigma_{LO}(Q^2=\tau s)  
\end{eqnarray}
where $\tau_0=4M_\Delta^2/s$ with $s$ being the total hadronic
c.m.\,energy squared, and ${\cal L}^{q\bar q}$ denotes the $q\bar q$
parton luminosity.

The standard QCD corrections, with virtual gluon exchange, gluon
emission and quark emission, are identical to the corresponding
corrections to the Drell--Yan process \cite{DY}. They modify the lowest
order cross section in the following way
\begin{eqnarray}
\sigma & = & \sigma_{LO} + \Delta\sigma_{q\bar q} + \Delta\sigma_{qg}
\nonumber \\
\Delta\sigma_{q\bar q} & = & \frac{\alpha_s(\mu_R)}{\pi} \int_{\tau_0}^1
d\tau \sum_q \frac{d{\cal L}^{q\bar q}}{d\tau} \int_{\tau_0/\tau}^1
dz~\hat \sigma_{LO}(Q^2 = \tau z s)~\omega_{q\bar q}(z) \nonumber \\
\Delta\sigma_{qg} & = & \frac{\alpha_s(\mu_R)}{\pi} \int_{\tau_0}^1
d\tau \sum_{q,\bar q} \frac{d{\cal L}^{qg}}{d\tau} \int_{\tau_0/\tau}^1
dz~\hat \sigma_{LO}(Q^2 = \tau z s)~\omega_{qg}(z)
\end{eqnarray}
with the coefficient functions \cite{DY}
\begin{eqnarray}
\omega_{q\bar q}(z) & = & -P_{qq}(z) \log \frac{\mu_F^2}{\tau s}
+ \frac{4}{3}\left\{ \left[\frac{\pi^2}{3} -4\right]\delta(1-z) +
2(1+z^2)\left(\frac{\log(1-z)}{1-z}\right)_+ 
\right\} \nonumber \\
\omega_{qg}(z) & = & -\frac{1}{2} P_{qg}(z) \log \left(
\frac{\mu_F^2}{(1-z)^2 \tau s} \right) + \frac{1}{8}\left\{ 1+6z-7z^2
\right\}
\end{eqnarray}
where $\mu_F$ denotes the factorization scale, $\mu_R$ the
renormalization scale and $P_{qq}, P_{qg}$  the well-known
DGLAP splitting functions, which are given by \cite{apsplit}
\begin{eqnarray}
P_{qq}(z) & = & \frac{4}{3} \left\{ \frac{1+z^2}{(1-z)_+}+\frac{3}{2}\delta(1-z)
\right\} \nonumber \\
P_{qg}(z) & = & \frac{1}{2} \left\{ z^2 + (1-z)^2 \right\} \, .
\end{eqnarray}

\section{Numerical Results} 
\begin{figure}[hbtp]
 \setlength{\unitlength}{1cm}
 \centering
 \begin{picture}(15,8.0)
  \put(2.0,-3.5){\epsfxsize=10cm \epsfbox{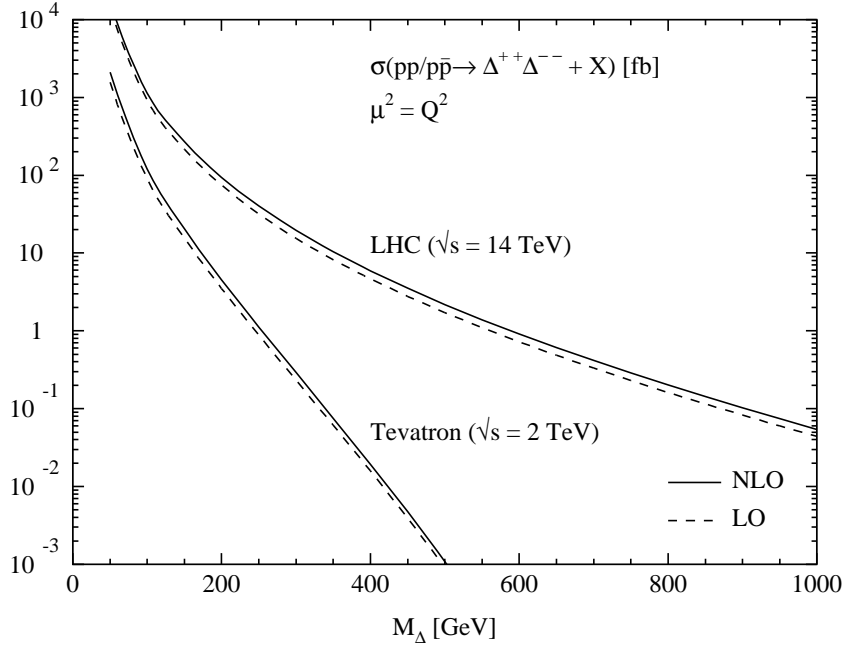}}
 \end{picture}
\caption[ ]{\it \label{fg:cxn} Production cross sections of
doubly-charged Higgs pair production at the Tevatron and the LHC. The
doubly-charged Higgs bosons $\Delta^{--}$ carry $I_{3\Delta}=-1$ as the
third isospin component. CTEQ6L1 (CTEQ6M) parton densities \cite{cteq6}
have been used at LO (NLO).}
\end{figure}

\begin{figure}[hbtp]
 \setlength{\unitlength}{1cm}
 \centering
 \begin{picture}(15,8.0)
  \put(2.0,-3.5){\epsfxsize=10cm \epsfbox{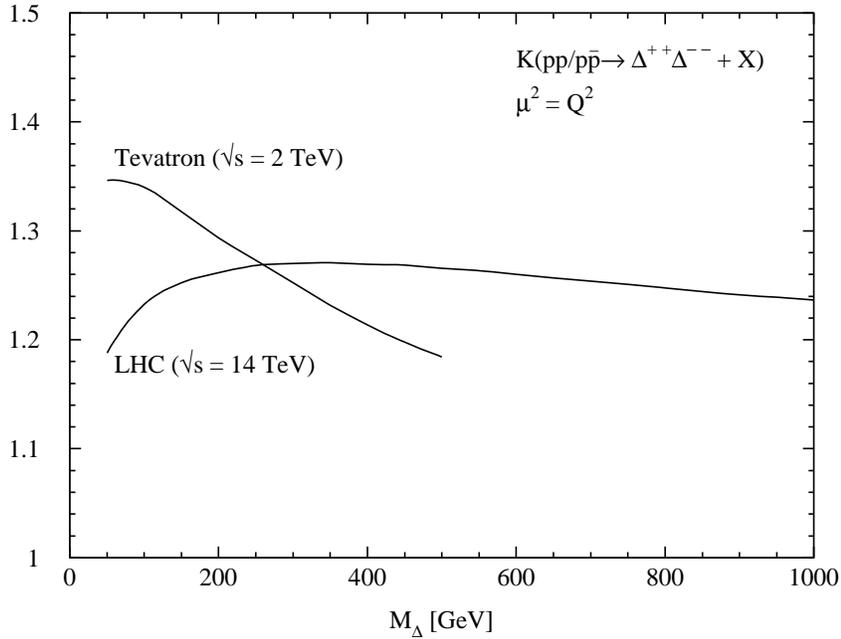}}
 \end{picture}
\caption[]{\it \label{fg:kfac} $K$ factors of doubly-charged Higgs pair
production at the Tevatron and the LHC. The parameters are the same as
in Fig.\,\ref{fg:cxn}. The curve for the Tevatron has been truncated at
$M_\Delta=500$ GeV, because the cross section is too small above and
thus phenomenologically irrelevant.}
\end{figure}

%        =================
Our numerical results will be presented using CTEQ6L1 (CTEQ6M) parton
densities \cite{cteq6} at (next-to-)leading order with the strong
coupling $\alpha_s$ adjusted accordingly, i.e.
$\alpha_s^{LO}(M_Z)=0.130, \alpha_s^{NLO}(M_Z)=0.118$.  The electroweak
quantum numbers of the doubly-charged Higgs boson $\Delta^{--}$ have
been chosen to be $I_{3\Delta}=-1$ and $e_\Delta=-2$.
Fig.\,\ref{fg:cxn} shows the total cross sections at the LHC and the
Tevatron in leading and next-to-leading order as a function of the
charged Higgs mass $M_\Delta$. The renormalization/factorization scale
has been chosen as $\mu_F^2=\mu_R^2=Q^2$ which is the natural scale
choice for Drell--Yan like processes. The QCD corrections increase the
cross sections by 20--30\% and are thus of moderate size. This can
explicitely be inferred from Fig.\,\ref{fg:kfac} where the $K$ factors,
defined as the ratio $K=\sigma_{NLO}/\sigma_{LO}$, are depicted for the
Tevatron and the LHC.  The curve for the Tevatron is truncated at
$M_\Delta=500$ GeV, since the cross section becomes too small above.
The residual renormalization and factorization scale dependence at NLO
amounts to about 5--10\% and serves as an estimate of the theoretical
uncertainties. They are of the order of the known NNLO corrections
\cite{nnlo} which amount to about 5--10\%. They have not been included in
our analysis. The uncertainties of the parton densities have to be added
resulting in a total theoretical uncertainty of about 10--15\%.

\section{Conclusions} 
%        ===========
In this note we have analyzed doubly-charged Higgs pair production at
the Tevatron and the LHC at NLO QCD. The NLO corrections increase the
cross sections by about 20--30\% and reduce the residual
renormalization/factorization scale dependence to 5--10\%. The total
theoretical uncertainties including the errors of the parton densities
can be estimated to be 10--15\%. This accuracy is sufficient for
doubly-charged Higgs boson searches at the Tevatron and LHC. \\

\noindent
{\bf Acknowledgements.} \\
We are grateful to S.\,Lammel for drawing our attention to this topic.
We thank P.M.\,Zerwas for carefully reading the manuscript.

\end{document}